\documentclass[aps,pra,twocolumn,superscriptaddress,showpacs,preprintnumbers,amsmath,amssymb]{revtex4-2}

\usepackage{newtxtext} 
\usepackage{newtxmath} 
\usepackage{graphicx}  
\usepackage{hyperref}  
\begin{document}
	
	
	\title{The new generation lunar gravitational wave detectors: sky map resolution and joint analysis}
	
	\author{Xiaolin Zhang}
	\author{Chengye Yu}
	\author{Haoran Li}
	\author{Sobhan Kazempour}
	\affiliation{School of Physics, Beijing Institute of Technology, Beijing, 100081, China}
	
	\author{Mingqiu Li}
	\email{limingqiu17@mails.ucas.ac.cn} 
	\affiliation{School of Physics, Beijing Institute of Technology, Beijing, 100081, China}
	
	\author{Sichun Sun}
	\email{sichunsun@gmail.com}
	\affiliation{School of Physics, Beijing Institute of Technology, Beijing, 100081, China}
	
	\date{\today} 
	
	\begin{abstract}
		Lunar-based gravitational-wave interferometry is a fascinating endeavor, and was proposed as a promising approach to bridge the observational gap between space- and ground-based detectors. In this work, we adopt the Fisher-matrix method to examine the angular-resolution performance of the newly proposed Crater Interferometry Gravitational-wave Observatory (CIGO) on the lunar crater rim near the north pole, together with TianQin and LISA, for monochromatic sources in the 0.1–10Hz band. We find that above 2.87 Hz, CIGO achieves better localization accuracy than the other two space-based missions and dominates the combined detector network's performance. We further explore an upgraded Tetrahedron configuration, TCIGO, with a fourth station at the bottom of a crater, which forms a regular tetrahedral constellation on the lunar surface. The result shows that TCIGO yields a five-fold improvement in angular-resolution capability over CIGO and gets better sky coverage across the target frequency band.
	\end{abstract}
	
	\maketitle 
	
	
	\section{Introduction}
	Gravitational-wave (GW) detection has become a crucial method of probing the nature of compact astrophysical objects under extreme conditions. Gravitational waves cover a broad frequency range from 
	$10^{-16}$ -- $10^{4}\,\mathrm{Hz}$ \cite{gong2021concepts,bailes2021gravitational,cai2024networks}.
	Existing GW detection missions are designed to probe different segments of this spectrum. Ground-based detectors, such as LIGO and Virgo \cite{abbott2016gw151226,abbott2017gw170104,abbott2017gw170814,abbott2018gw170817,abbott2017gw170817,abbott2019gwtc,abbott2009ligo,abbott2016observation}, are sensitive to frequencies in the range of 
	$3$ -- $2\times10^{4}\,\mathrm{Hz}$, 
	while space-based interferometers like LISA  \cite{amaro2017laser,ESA_LISA_2025} TianQin \cite{luo2016tianqin,luo2025progress} and Taiji \cite{hu2017taiji,luo2021taiji}target the lower-frequency band of 
	$10^{-4}$ -- $10^{-1}\,\mathrm{Hz}$.
	However, the intermediate deci-hertz range ($0.1$–$10\,\mathrm{Hz}$) lies outside the optimal sensitivity bands of both ground- and space-based detectors.
	
	The deci-hertz window ($0.1$--$10\,\mathrm{Hz}$) is of particular astrophysical interest, as it encompasses numerous rich phenomena \cite{kuroda2015gravitational,gao2021mid}, such as the mergers of intermediate-mass black holes, compact binary white dwarf systems, and core-collapse supernovae \cite{sousa2025parameter,jiang2024sky,seitenzahl2015neutrino}.
	Moreover, it provides a unique opportunity to test General Relativity \cite{sedda2020missing,jani2020detectability}and extensions of the Standard Model \cite{yao2025axion,xu2025distinguishing,wang2023frequency,figliolia2026gravitational}.
	Thus, constructing detectors optimized for this frequency range is essential for probing these phenomena.
	
	Several proposals have been put forward for \textbf{space-based} deci-hertz GW observatories to bridge this gap,  including the geocentric mission like SAGE \cite{sousa2025parameter} 
	and heliocentric concepts such as DECIGO \cite{kawamura2019space,
		kawamura2021current}, TVLBAI \cite{abend2024terrestrial} and ALIA \cite{crowder2005beyond}, and for \textbf{earth-based} detectors including TOBA \cite{shimoda2020torsion}, AION \cite{badurina2020aion}and ZAIGA \cite{zhan2020zaiga}.
	Recently, lunar-based GW detection has attracted increasing attention, aiming to fill this observational blind spot \cite{abend2024terrestrial,baker2019space,kawamura2019space,branchesi2023lunar,lafave1993lunar,cozzumbo2024opportunities,song2025probing,yan2024toward}.
	The \textbf{lunar} GW observatory concept envisions the construction of an interferometric detector on the Moon. 
	GLOC (Gravitational-Wave Lunar Observatory for Cosmology) \cite{jani2021gravitational} and LILA (Laser interferometer lunar antenna) \cite{jani2025laser} missions were proposed, aiming to build a 40-km equilateral triangle laser interferometer on the lunar surface as part of NASA’s Artemis program. 
	LGWA (lunar gravitational-wave antenna) \cite{harms2021lunar,ajith2025lunar,song2025probing} deploys an array of high-end seismometers on the Moon to monitor its normal modes in the $1\,\mathrm{mHz}$–$1\,\mathrm{Hz}$ band excited by GWs. 
	Additionally, Italian researchers proposed the LSGA (Lunar Seismic and Gravitational Antenna) \cite{katsanevas2020ideas} project to detect GW-induced lunar seismic strains.
	
	There are a few advantages to constructing GW detectors on the Moon. 
	Compared with \textbf{earth-based} detectors, firstly, the Moon naturally provides an ultra-high-vacuum environment, with surface pressures as low as $10^{-8}\,\mathrm{Pa}$ at sunrise, comparable to the LIGO vacuum system, significantly reducing construction and maintenance costs \cite{johnson1972vacuum}. Secondly, at low frequencies ($0.1$–$0.4\,\mathrm{Hz}$), lunar seismic noise is three to four orders of magnitude lower than that on Earth \cite{harms2022seismic,jani2021gravitational}.
	Thirdly, the absence of atmospheric, human, and tectonic disturbances allows for stable, long-term operation with a high duty cycle \cite{jani2025laser,de2007modeling,horanyi2015permanent}. Consequently, gravity-gradient noise is also strongly suppressed, thereby minimizing the need for costly mitigation strategies \cite{abend2024terrestrial,branchesi2023lunar,lafave1993lunar,cozzumbo2024opportunities,song2025probing,yan2024toward}. Compared with \textbf{space-based} detectors, a lunar-based GW observatory, stationed on the surface, can be serviced directly by astronauts, resulting in far lower maintenance costs than those of deep space missions \cite{niu2024lunar}. Secondly, for low-frequency signals around $0.1\,\mathrm{Hz}$, the data transmission rate required for a lunar observatory is only about one order of magnitude higher than that of space-based missions, making data transfer considerably more manageable \cite{kawamura2019space,baker2019space,larose2005lunar,lognonne2009moon,niu2024lunar}. Together, these advantages make the Moon an auspicious site for next-generation gravitational-wave astronomy, offering a stable, low-noise, and serviceable platform that bridges the spectrum gap between ground and space-based detectors.
	
	Several leading research institutions in China, including the Beijing Institute of Technology and the Chinese Academy of Sciences, have jointly proposed a lunar-based gravitational-wave (GW) detection concept known as the Crater Interferometry Gravitational wave Observatory (CIGO) \cite{CIGO_LZU_2024,CIGO_COSE_2024}. CIGO is conceived as a third-generation laser interferometric detector deployed on the lunar surface, aiming at a target strain sensitivity of $\sim 10^{-24}$, approximately one order of magnitude beyond current second-generation ground-based detectors ($\sim 10^{-23}$). The instrument is designed to operate in the 0.1–10 Hz frequency band, bridging the observational gap between space- and earth-based missions.
	
	Unlike space-based missions such as LISA and TianQin, CIGO is not a free-flying constellation. Instead, it consists of three end stations rigidly anchored to the lunar surface, arranged in an equilateral triangular configuration near the lunar north pole (see Fig. 8). The nominal arm length adopted in this work is 100 km. The constellation rotates with the Moon. 
	The detector concept assumes a standard equal-arm Michelson interferometric configuration with stabilized laser links established between surface stations. Owing to its rigid geometry and equal-arm design, time-delay interferometry (TDI), which is essential for space-based unequal-arm interferometers, is not required. The interferometric response is therefore described by the standard finite-arm laser interferometer transfer function derived from the light round-trip time delay in the presence of gravitational waves, which is directly applicable to the lunar-based configuration considered here.
	
	In this work, we focus on the monochromatic gravitational waves, which can be emitted in the early stage of compact binary systems, including intermediate-mass black holes and neutron stars \cite{seitenzahl2015neutrino,sousa2025parameter,jiang2024sky,song2025probing}, the first-order phase transitions
	from the early universe \cite{croon2024gravitational} and the subsequent cosmic string \cite{zhou2022gravitational}. As with space-based detectors, we adopt an equilateral triangular configuration (see Fig. 8). Using the \textit{Fisher Information Matrix (FIM)} method, we analyze $3600$ potential GW sources distributed across the celestial sphere to evaluate the localization accuracy of the detector. We further generate sky maps of angular resolution for CIGO, LISA, and TianQin at three representative frequencies: $0.1$, $1$, and $10~\mathrm{Hz}$. By comparing these results, we investigate the performance of the detectors across the $0.1$--$10~\mathrm{Hz}$ Hz band. Since the operation of Taiji in space is similar to LISA's, we will not discuss it separately here. 
	Besides, we compute joint angular-resolution maps for the CIGO, LISA, and TianQin network configurations at the same frequencies, and assess the enhanced localization capability using a lunar–space joint analysis. Additionally, we upgrade the CIGO system by adding a detector at the base of the crater, forming a tetrahedral configuration with the origin plane. We then calculate and compare its angular resolution over the sky region. These results provide valuable insight into the potential of lunar-based interferometers to bridge the existing detection gap between space-based and ground-based GW observatories.
	
	The paper is organized as follows. In the results section, we present the sky-localization performance of CIGO operating both independently and in a network with LISA and TianQin, and estimate the impact of lunar-based noise on CIGO's localization accuracy. Besides, we investigate the regular tetrahedral configuration of CIGO and compare its angular resolution with that of the original planar configuration to assess potential improvements. We conclude all the results and provide an outlook in the discussion section. In the methods section, we outline the methodology and settings adopted in this work, including the Fisher information matrix formalism for monochromatic gravitational waves and the orbital equation of CIGO detectors. Some detailed derivations and parameter settings are presented in the supplementary information.
	
	\section*{Results}

	\subsection*{Source Localization}

	The Crater Interferometry Gravitational-wave Observatory (CIGO) is a proposed equilateral-triangle laser interferometer designed to probe mid-frequency gravitational waves in the deci-hertz to 10-hertz range. Each arm has a nominal length of approximately $10^2$ km, aiming for a target strain sensitivity of $\sim 10^{-24}$.
	
	We evaluate the angular-localization accuracy of CIGO and compare it with two representative space-based detectors to assess its performance. LISA and TianQin, which adopt similar triangular configurations. LISA consists of three spacecraft forming a triangle with $2.5 \times 10^6$ km arm lengths, trailing Earth by about 20° in the ecliptic plane with an orbital plane inclined at 60°. While TianQin has arm lengths of $1.73 \times 10^5$ km and its normal vector of the detector plane points to RX J0806.3+1527 with the latitude $\theta_{tq} = -4.7^\circ$ and the longitude $\phi_{tq} = 120.5^\circ$ in heliocentric coordinates. Both detectors operate in an Earth-centered orbit.
	
	The distinct orbital dynamics and orientations of these missions introduce different levels of amplitude and Doppler modulation in their detector responses \cite{cutler1998angular,blaut2011accuracy}, which, in turn, strongly influence their sky-localization capabilities. By comparing the sky maps of angular resolution across these three detectors in the mid-frequency band, we can clearly identify the optimal operating frequency range and relative advantages of CIGO.
	
	In this section, we adopt the FIM method described above, simulate 3600 gravitational-wave sources uniformly distributed across the sky, and compute angular-resolution sky maps for CIGO, LISA, and TianQin. We set the mission time to one year and the observation coverage to $-\pi/2 < \theta_s < \pi/2$ in latitude and $-\pi < \phi_s < \pi$ in longitude. The gravitational-wave signals are modeled as monochromatic waves, characterized by a simple waveform and a minimal set of parameters. To ensure the detectability of all signal sources, we set the SNR threshold for all sources to 7.
	
	\begin{figure*}[t] 
		\centering
		\includegraphics[width=0.95\textwidth]{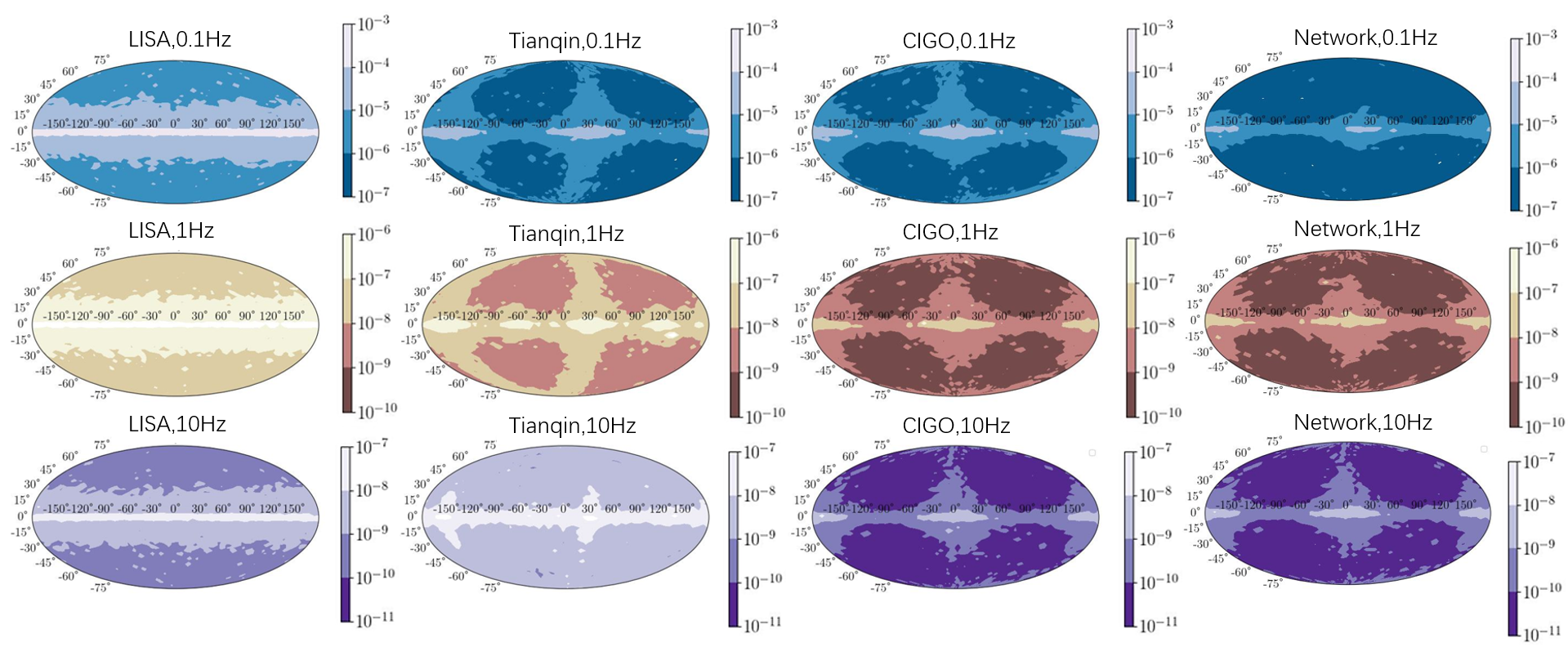} 
		
		\caption{The sky map of angular resolutions $\Delta \Omega_S$ of sources from different directions in the unit of steradian (1 steradian is 3000 square degrees) for LISA, Tianqin, CIGO, and their network. The horizontal axis represents the longitude $\phi_s$ and the vertical axis represents the latitude $\theta_s$. The frequencies of monochromatic sources are $0.1$ Hz, $1$ Hz, and $10$ Hz.} 
		\label{fig:1}
	\end{figure*}
	
		\begin{figure*}[t] 
		\centering
		\includegraphics[width=0.95\textwidth]{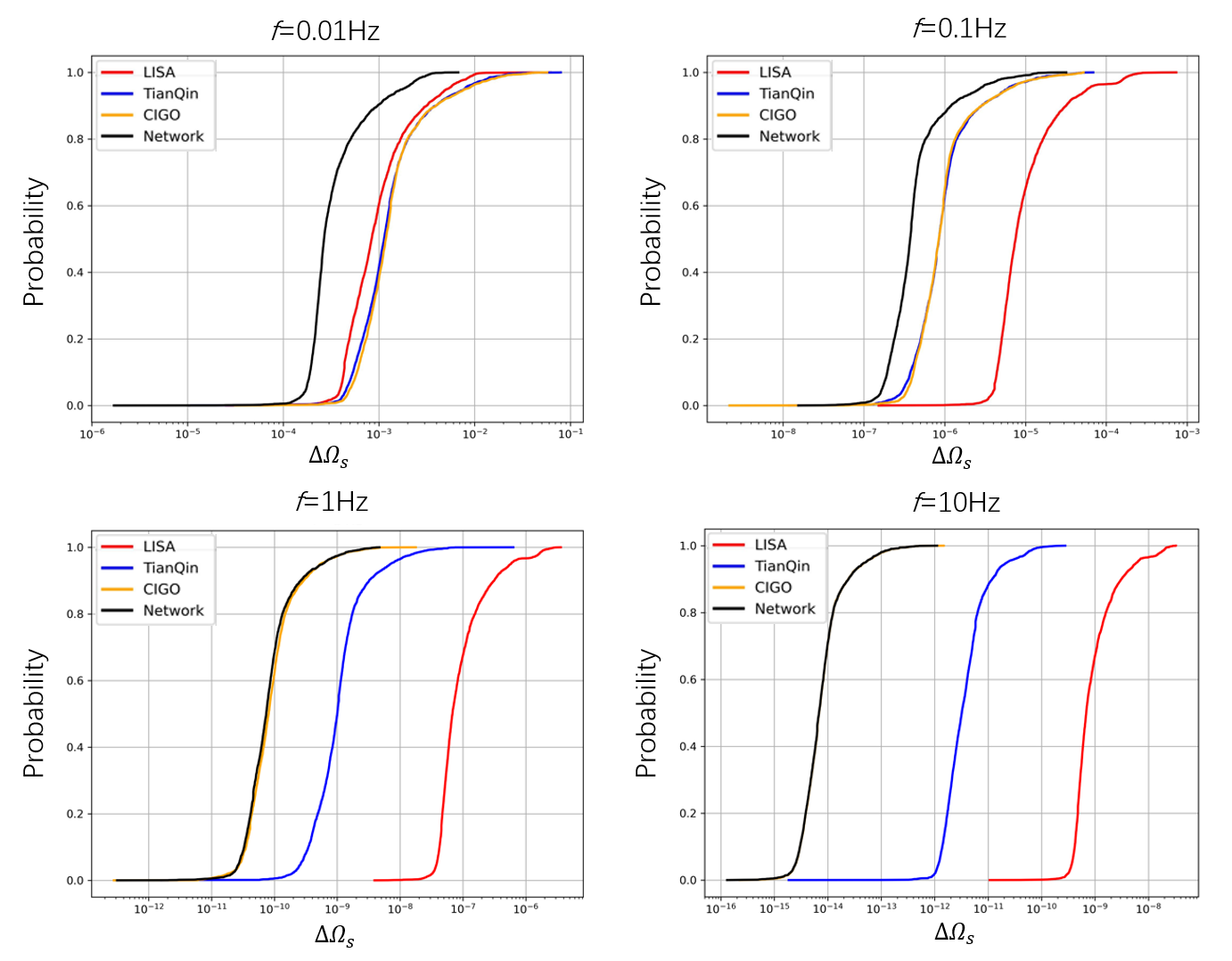} 
		
		\caption{Cumulative histograms of sky localization estimations $\Delta \Omega_S$ for  LISA (red), TianQin (blue), CIGO (yellow), and their joint network (black). From left to right and top to bottom the panels correspond to representative frequencies $f = 0.01\,\mathrm{Hz}$, $0.1\,\mathrm{Hz}$, $1\,\mathrm{Hz}$, and $10\,\mathrm{Hz}$. Smaller values of $\Delta \Omega_s$ indicate better angular resolution.} 
		\label{fig:2}
	\end{figure*}
	
	The results of angular resolutions at the frequencies $f = 10^{-1}$ Hz, ${1}$ Hz and $10$ Hz for the LISA, Tianqin and CIGO are shown in Fig.1 and Fig.2. We show the sky map of the angular resolution for  LISA, Tianqin and CIGO in Fig.1, and the cumulative histograms of the angular resolution in Fig.2. The mean values of angular resolutions are summarized in Table~I.

	Fig.1 presents the sky maps of angular resolution for 3600 monochromatic gravitational-wave sources uniformly distributed over the celestial sphere, computed for LISA, TianQin, CIGO, and their network at frequencies of 0.1~Hz, 1~Hz, and 10~Hz. The horizontal and vertical axes correspond to the source longitude ($-180^\circ$ to $180^\circ$) and latitude ($-90^\circ$ to $90^\circ$), respectively. The color scale represents the localization uncertainty $\Delta\Omega_s$ in steradians (deg), where darker regions indicate better (smaller) angular resolution and lighter areas correspond to poorer performance. From top to bottom, the rows correspond to increasing frequencies (0.1~Hz, 1~Hz, and 10~Hz), while each column corresponds to a specific detector: LISA (left), TianQin (middle), and CIGO (right).

	For LISA, we find that at all frequencies, the localization accuracy $\Delta\Omega_s$ for sources near the ecliptic plane is typically two orders of magnitude worse than that for sources near the poles. In contrast, TianQin shows degraded angular resolution for sources located along its detector plane (around $\phi_s \approx -150^\circ$ and $30^\circ$) and near the heliocentric equatorial plane, where $\Delta\Omega_s$ is roughly two orders of magnitude larger than in other sky regions. Similarly, for CIGO, the localization accuracy deteriorates by about one order of magnitude for sources located near the longitude $\phi_s = 5^\circ$ and $180^\circ$, as well as near the lunar equatorial plane.
	
	These patterns arise from the differences in the orbital geometry and orientation of the detector planes. TianQin’s detector usually continuously points toward its reference source RX~J0806.3+1527, located at latitude $\theta = -4.7^\circ$ and longitude $\phi = 120.5^\circ$. CIGO, in contrast, is fixed on the lunar surface near the north pole, where the lunar spin axis is nearly parallel to the normal of the ecliptic plane, and the inclination between the lunar equator and the ecliptic (the so-called white-plane inclination) is about $5^\circ$. Consequently, CIGO’s detector plane faces approximately toward $\theta = 5^\circ$. Along the detector-plane normal vector, the signal-to-noise ratio is significantly reduced, leading to reduced localization accuracy (larger $\Delta\Omega_s$) and weaker modulation \cite{zhang2021sky}. LISA, however, lacks a fixed reference source; its detector plane rotates around the Sun with an inclination of $60^\circ$, inducing strong amplitude modulations that significantly enhance sky-localization performance across most directions. As a result, LISA exhibits a more uniform angular resolution over the sky map. At the same time, TianQin and CIGO show slightly less sky coverage, characterized by high precision over large regions but weaker performance in specific, limited directions.
	
	One point that needs emphasis is the impact of uncertainties in lunar orbital and rotational modeling. Actually, in contrast to heliocentric free-flying constellations,
	CIGO is rigidly anchored on the lunar surface. Therefore, the detector geometry is fixed in the
	lunar body frame, and the only time-dependent modulation relevant to sky localization arises
	from the rotation and precession of the Moon, which will induce the deflection of the detector plane.
	Based on the residual levels reported in the JPL ephemerides \cite{park2021jpl}, we adopt a very conservative uncertainty in the lunar libration angles,
	$\delta\theta_c \approx 5~\mathrm mas $, such an perturbation would induce a shift of angular resolution $\delta(\Delta\Omega_s)$, and the resulting fractional change is
	\begin{equation}
		\frac{\delta(\Delta\Omega_s)}{\Delta\Omega_s}
		\lesssim 10^{-7}.
	\end{equation}
	
	This level of shift is entirely negligible compared with the intrinsic Fisher-matrix
	uncertainties and instrumental noise contributions.
	
	Importantly, small perturbations in the lunar orientation do not change the overall scale
	of the angular resolution. Instead, they induce only a tiny redistribution of the sky-location
	error pattern. In other words, the global angular resolution capability of the detector
	remains unaffected, while the directional dependence of the error ellipse may experience
	an imperceptible shift at the level of $10^{-7}$. 
	
	Fig.2 illustrates the cumulative distribution functions (CDFs) of the angular resolution
	$\Delta\Omega_s$ for the same set of sources across different frequencies, with curves for 
	LISA (red), TianQin (blue), CIGO (yellow), and their combined network (black). 
	The $x$-axis is logarithmic in $\Delta\Omega_s$ (deg), and the $y$-axis shows the cumulative 
	probability, i.e., the fraction of sources localized below a given uncertainty. 
	The panels are arranged by frequency: left for $0.1\,\mathrm{Hz}$, middle for $1\,\mathrm{Hz}$,
	and right for $10\,\mathrm{Hz}$.
	
	At $f = 0.1\,\mathrm{Hz}$, both CIGO and TianQin exhibit localization accuracies of 
	$\Delta\Omega_s \sim 10^{-6}\,\mathrm{deg}$, whereas LISA’s resolution is reduced by roughly one 
	order of magnitude. As the frequency increases to $10\,\mathrm{Hz}$, CIGO’s localization 
	performance surpasses that of the other two detectors by more than two orders of magnitude. However, in our target band (0.1--10\, Hz), the observing frequency satisfies $f \ll f^* = c/(2\pi L)$, the angular resolution $\Delta \Omega_s$ of a single detector improves with increasing gravitational-wave frequency. 
	In contrast, for LISA ($f^* \simeq 0.02$\, Hz) and TianQin ($f^* \simeq 0.28$\, Hz), the target band lies well above their frequencies. Consequently, the transfer function's frequency dependence deteriorates markedly, causing the responses of LISA and TianQin to saturate.
	
	For the network configuration combining CIGO, TianQin, and LISA, the joint observations 
	provide a clear improvement at low frequencies ($f = 0.01$--$0.1\,\mathrm{Hz}$) when compared 
	to any single detector. In this regime, the decreasing regions of sky map are well-complementary, and the cumulative distributions of $\Delta\Omega_s$ shift 
	leftward relative to the individual-detector curves, reflecting the benefit of the geometric 
	complementarity among the three non-coplanar constellations.
	
	As the frequency increases into the $1$--$10\,\mathrm{Hz}$ range, however, the network’s 
	performance becomes overwhelmingly dominated by CIGO. As indicated by the black CDF curve in 
	Fig. 2, the network curve coincides with that of CIGO at these frequencies, showing 
	that CIGO’s response effectively sets the localization accuracy of the joint observation. 
	In this high-frequency regime, the contributions from LISA and TianQin become negligible, and 
	the network inherits CIGO’s superior angular-resolution capability.
	
	\begin{table}[!ht]
		\centering
		
		\renewcommand{\arraystretch}{1.3}
		\begin{tabular}{|c|c|c|c|c|}
			\hline
			\boldmath $f$ (Hz) 
			& \textbf{LISA} 
			& \textbf{TianQin} 
			& \textbf{CIGO} 
			& \textbf{Network} \\
			\hline
			& \multicolumn{4}{c|}{Mean value of angular resolution} \\
			\hline
			$0.01$ 
			& $1.874 \times 10^{-3}$ 
			& $2.633 \times 10^{-3}$ 
			& $2.845 \times 10^{-3}$ 
			& $5.783 \times 10^{-4}$ \\
			\hline
			$0.1$ 
			& $2.376 \times 10^{-5}$ 
			& $2.250 \times 10^{-6}$ 
			& $2.296 \times 10^{-6}$ 
			& $9.221 \times 10^{-7}$ \\
			\hline
			$1$ 
			& $2.266 \times 10^{-7}$ 
			& $2.930 \times 10^{-9}$ 
			& $2.272 \times 10^{-10}$ 
			& $1.967 \times 10^{-10}$ \\
			\hline
			$10$ 
			& $2.350 \times 10^{-9}$ 
			& $8.719 \times 10^{-12}$ 
			& $1.969 \times 10^{-14}$ 
			& $1.927 \times 10^{-14}$ \\
			\hline
		\end{tabular}
		\caption{Angular resolution (in steradians) comparison for
			LISA, TianQin, CIGO, and their combined network at different
			frequencies}
		\label{tab:resolution_comparison}
	\end{table}
	
	For Michelson interferometers operating in space, such as TianQin and LISA, the analytical form of the single-link noise power spectral density is written as \cite{robson2019construction,babak2021lisa,cornish2001detecting}
	\begin{equation}
		S_n(f) = \frac{S_x}{L^2}
		+ \frac{4 S_a}{(2\pi f)^4 L^2}
		\left( 1 + \frac{10^{-4}\,\mathrm{Hz}}{f} \right),
	\end{equation}
	where $L$ denotes the interferometer arm length, $S_x^{1/2}$ and $S_a^{1/2}$ are displacement noise and acceleration noise, respectively.
	Since there is currently no formula for estimating quantum noise for a CIGO-like detector, which is a Michelson-laser interferometer operating on a planet, in this work, we first naively assume that the noise level of CIGO can be achieved at least at the noise level in TianQin \cite{zhang2021sky} with the same frequency dependence. 
	
	One notable feature is that the CIGO detector is designed as a Michelson-laser interferometer deployed on the lunar surface. If we examine the detector's noise budget more closely, CIGO's performance can vary significantly with noise level. Notice that the formalism of Michelson-laser power spectral density is largely implementation-agnostic, so it's also suitable for CIGO \cite{sathyaprakash2009physics,moore2014gravitational}. Here, we estimate the noise based on the ground-based program. Given that CIGO is deployed on the moon, it will be affected by the lunar environment, to assess for these effects, we employ the \texttt{pygwinc} package \cite{pygwinc2020}and assume that CIGO adopts detector technologies comparable to those of the Cosmic Explorer Stage 1 (CE1) \cite{hall2022cosmic}, including the test masses, thermal shielding, and vibration isolation systems,\textit{etc.} Starting from the CE1 design noise budget, we modify the environmental parameters corespond to lunar conditions and obtain the corresponding lunar environmental strain noise.
	Following Ref \cite{moore2014gravitational,sathyaprakash2009physics}, we adopt the relation below to connect the noise power spectral density $S_n(f)$ and the strain $h_n(f)$:
	\begin{equation}
		S_n(f) = \left| h_n(f) \right|^2
	\end{equation}
	Substituting into equation (17), we obtain the overall strain-noise budget, shown in Fig. 3. For instructions on setting and modifying the relevant parameters, please refer to Supplementary  Information B.
	

	
	\begin{figure}[htbp] 
		\centering
		\includegraphics[width=0.95\linewidth]{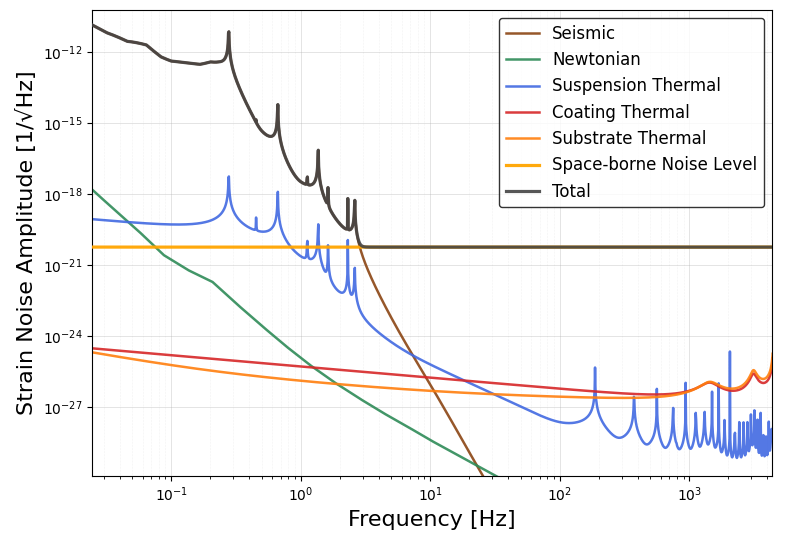} 
		
		\caption{The lunar environmental noise level estimation of CIGO. The strain noise amplitude spectral density (in $\mathrm{Hz}^{-1/2}$) is shown as a function of frequency for the dominant lunar noise sources: seismic (brown), Newtonian (green), suspension thermal (blue), coating thermal (red), substrate thermal (orange), and the space-borne noise level (yellow horizontal line). Total noise (black) is the square root of the sum of the squares of all contributions.} 
		\label{fig:3} 
	\end{figure}
	As shown in Fig. 3, within the target frequency band (0.1-10Hz) of CIGO, if we assume that the overall sensitivity level of CIGO is comparable to that of CE1, then even though the lunar seismic environment is quieter than that on Earth, the total noise curve is still significantly elevated in the frequency range below 2.87 Hz due to seismic noise contributions. Applying power-law fitting and smoothing procedure \cite{jani2021gravitational,cornish2001space,bender1997confusion}, we obtain an analytical expression for the noise strain amplitude for CIGO below:
	\begin{equation}
		h(f) =
		\begin{cases}
			5.57 \times 10^{-15}\, f^{-2.15}, 
			& 0.01 < f < 0.25\,\mathrm{Hz}, \\[4pt]
			1.12 \times 10^{-17}\, f^{-7.08}, 
			& 2.15 < f < 2.87\,\mathrm{Hz}, \\[4pt]
			\sqrt{S_n(f)}, 
			& f>2.87Hz.
		\end{cases}
	\end{equation}
	
	By replacing the $S_n(f)$ in Eq. (15) with Eq. (21), we recompute the angular resolution of CIGO, and the cumulative distribution histograms of the angular resolution are shown in Fig. 4.

	\begin{figure*}[t] 
		\centering
		\includegraphics[width=0.95\textwidth]{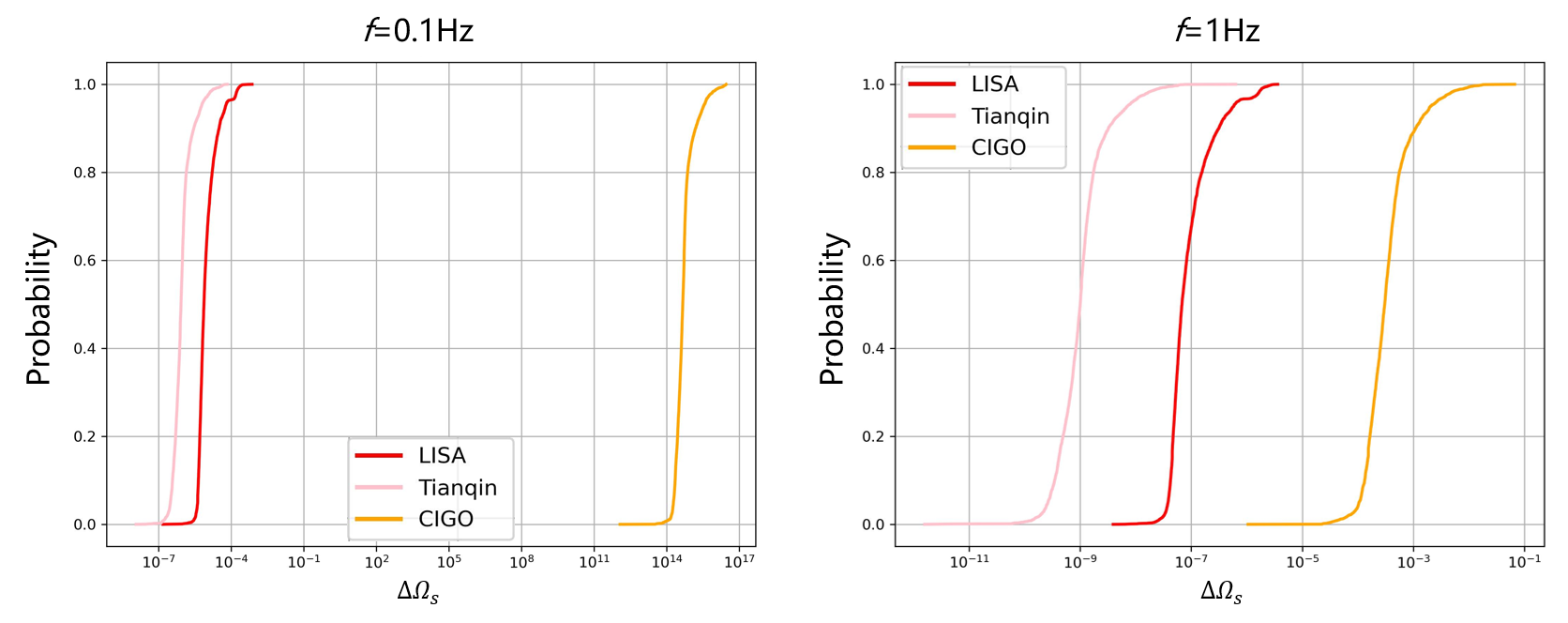} 
		
		\caption{Cumulative histograms of sky localization estimations $\Delta \Omega_S$ considering the lunar noise. From left to right and top to bottom the panels correspond to representative frequencies $f = 0.01\,\mathrm{Hz}$, $0.1\,\mathrm{Hz}$, $1\,\mathrm{Hz}$, and $10\,\mathrm{Hz}$. Smaller values of $\Delta \Omega_s$ indicate better angular resolution.} 
		\label{fig:4}
	\end{figure*}
	
	Fig. 4 shows the cumulative histograms of the angular resolution $\Delta\Omega_s$ at 0.1 and 1Hz after accounting for lunar environmental noise. It's seen that within the seismic-dominated frequency range of 0.1–2.87 Hz, the angular resolution of CIGO is substantially reduced. This indicates that, for a lunar-based implementation of CIGO, more advanced seismic isolation techniques and improved suspension thermal-noise suppression will be required to reduce instrumental noise further and achieve the localization performance demonstrated in Fig. 2. 
	
	Actually, such low-frequency seismic noise is a critical technical challenge for any lunar-based interferometer. However, several mitigation strategies may be considered. For example, it has been reported in the literature that by using the Suspension Platform Interferometer (SPI) technology, the length fluctuations of the detector in a suspended optical resonator can be reduced by three orders of magnitude \cite{koehlenbeck2023study,matichard2015seismic}. Additionally, installing environmental sensors, such as seismometers and microphones, can reduce gravitational noise \cite{harms2015newtonian}. Besides, our estimation of the noise spectrum is intentionally conservative. Compared to other posted proposals, such as GLOC \cite{jani2021gravitational}, our total noise level is more than three orders of magnitude higher in the relevant frequency band. This difference likely arises because GLOC adopts the default \texttt{pygwinc} quantum noise to estimate the shot noise induced by photon fluctuations in the laser beam, whereas we implement a space-based formalism. The precise noise curve will ultimately depend on the detailed detector configuration of CIGO and can be adjusted accordingly once the instrumental design is specified.
	
	It should be clarified that the above noise modeling is carried out under the assumptions of stationary Gaussian noise and thermal equilibrium. All modeling is conducted in the frequency domain and does not consider time-evolution characteristics.

	\begin{figure}[htbp] 
		\centering
		\includegraphics[width=0.95\linewidth]{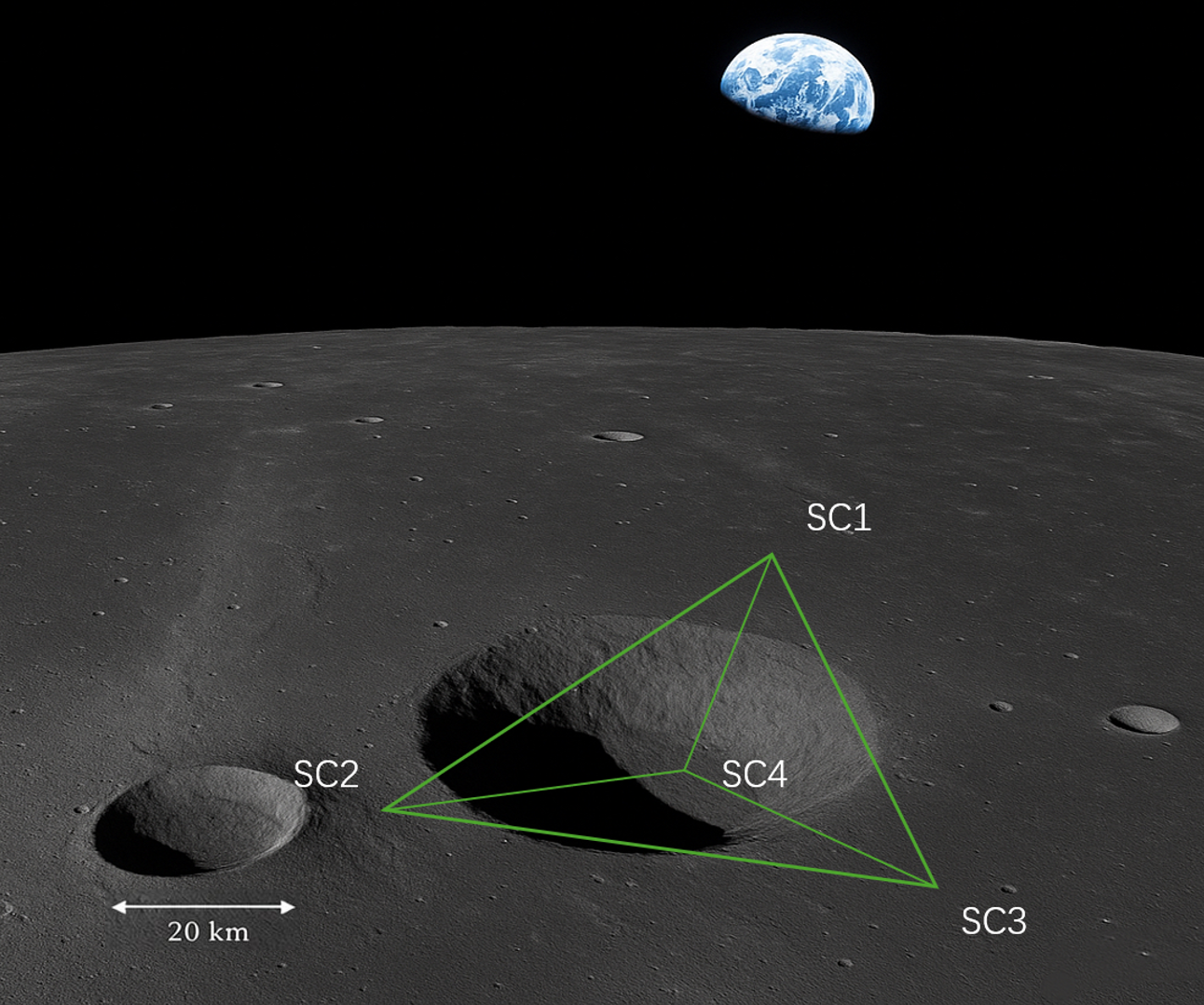} 
		
		\caption{\textbf{Conceptual design of TCIGO.}Three spacecraft (SC1--SC3) are situated on the rim of a lunar crater, with a fourth (SC4) located at its bottom, forming a regular tetrahedral configuration. Each end station, separated by 100 km, houses a test mass and three laser systems. Image of the lunar surface is adapted from Ref.~\cite{jani2021gravitational}.} 
		\label{fig:5}
	\end{figure}

	\subsection*{Tetrahedron Form of CIGO}
	In this section, we consider an upgraded configuration of the CIGO detector, referred to as the \textit{Tetrahedron Crater Interferometry Gravitational-wave Observatory} (TCIGO). As illustrated in Fig.~5, we augment the original lunar-polar CIGO by adding an interferometric station at the bottom of the crater. The four stations form a regular tetrahedral geometry. Each triangular face functions effectively as two independent laser interferometers, introducing three additional baselines with distinct orientations. Consequently, a single observation can be treated as an eight-interferometer network. This configuration is proposed to improve sky-localization performance for lunar-based detectors.
	
	We label the four interferometric stations as SC1--SC4. Following the orbital prescription of CIGO and adopting the settings of \cite{rubbo2004forward,jin2025tetrahedron}, the position functions of the \( n \)-th detector are identical to Eqs.~(10)--(13), the angle
	\( \alpha_n(t) = 2\pi f_n t + \kappa_n \)represents the angular displacement of the \( n \)-th detector in the detector-plane / lunar-polar coordinate system, with
	\( \kappa_n = 2\pi (n-1)/3 \), (n = 1, 2, 3, 3.5)characterizing the relative phase between SC1 and SC4. All other parameters follow the "Source Localization" section. $L_t$ is the distance from the center of the tetrahedron to each detector. For SC1--SC3, we use \( L_t= 40\,\mathrm{km} \), while for SC4, we take \( L_t= 57.16\,\mathrm{km} \), such that the four stations form a regular tetrahedron. 
	
	However, it is important to note that such a regular tetrahedral configuration is more of an idealized discussion. The purpose of introducing it is to study how adding the fourth station can improve the angular resolution and sky coverage of CIGO. Actually, such deep craters on the moon's surface are virtually non-existent. In practice, the non-equal-arm configuration may be preferable to the regular tetrahedron.
	
	We refer to the triangular detector formed by SC1--SC2--SC4 as \textit{S1} and that formed by SC1--SC3--SC4 as \textit{S2}. The original CIGO corresponds to the triangle SC1--SC2--SC3. Fig.~6 presents the sky maps of angular-resolution performance for 3600 monochromatic gravitational-wave sources uniformly distributed over the celestial sphere for S1 (left), CIGO (middle), and TCIGO (right), evaluated at frequencies of 0.1\, Hz, 1\, Hz, and 10\, Hz. Fig.~7 shows the cumulative distribution functions (CDFs) of the angular resolution \( \Delta\Omega_s \) for the same set of sources across all frequencies, with curves for LISA (red), S1 (blue), S2 (pink; nearly overlapping with S1), CIGO (yellow), and TCIGO (purple).

	\begin{figure*}[t] 
		\centering
		\includegraphics[width=0.95\textwidth]{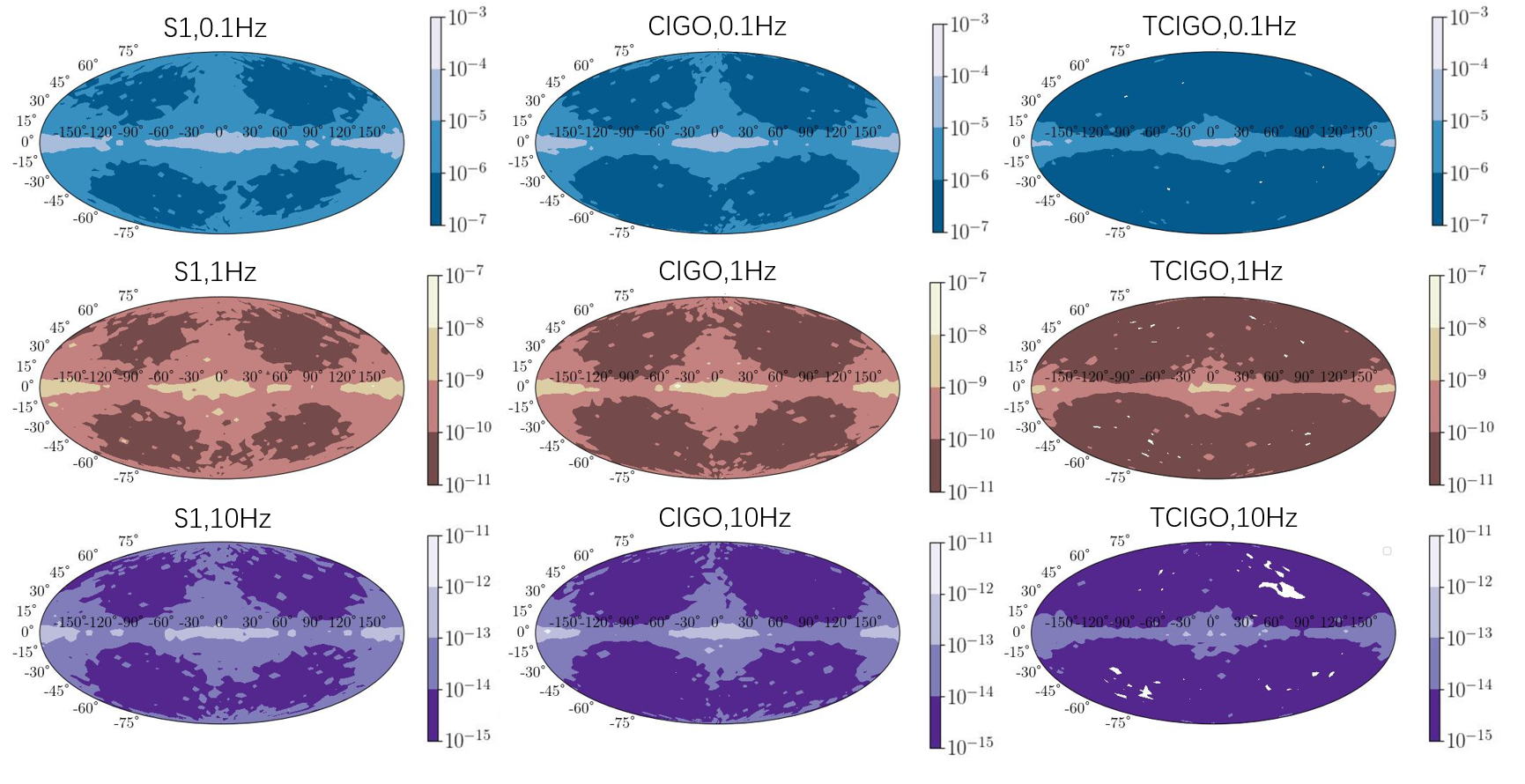} 
		
		\caption{The sky map of angular resolutions $\Delta \Omega_S$ of sources from different directions in the unit of steradian (1 steradian is 3000 square degrees) for S1, CIGO, and TCIGO. The horizontal axis represents the longitude $\phi_s$ and the vertical axis represents the latitude $\theta_s$. The left panel shows S1, the middle panel shows CIGO, and the right panel shows TCIGO. From top to bottom, the frequencies of monochromatic sources are $0.1$ Hz, $1$ Hz, and $10$ Hz.} 
		\label{fig:6}
	\end{figure*}

	\begin{figure*}[t] 
		\centering
		\includegraphics[width=0.95\textwidth]{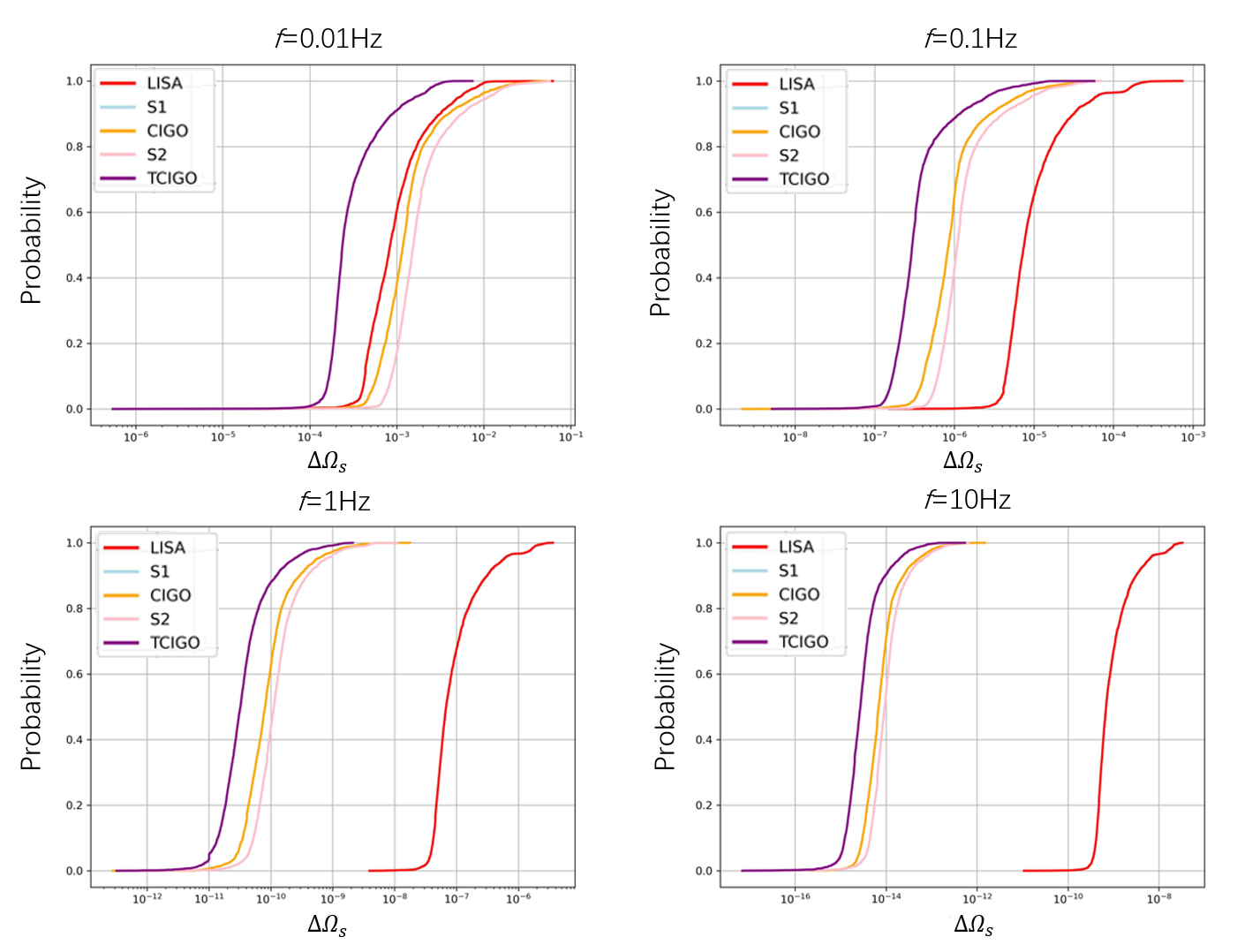} 
		
		\caption{Cumulative histograms of sky localization estimations $\Delta \Omega_S$ for LISA (red), TianQin (blue), CIGO (yellow) and TCIGO(purple) . From left to right and top to bottom the panels correspond to representative frequencies $f = 0.01\,\mathrm{Hz}$, $0.1\,\mathrm{Hz}$, $1\,\mathrm{Hz}$, and $10\,\mathrm{Hz}$. Smaller values of $\Delta \Omega_s$ indicate better angular resolution.} 
		\label{fig:7}
	\end{figure*}
	From Fig.~6, we observe that for a single detector formed with SC4 (e.g., S1), the sky-localization distribution and precision are comparable to those of CIGO. However, its high-precision region is slightly smaller. However, when all four faces are combined to form the full TCIGO array, the reduction in angular resolution caused by the orientation of the CIGO (latitude $\theta = 5^{°}$) was compensated. As seen in Fig.~7, at $ f = 0.1\,\mathrm{Hz} $, the cumulative localization performance of TCIGO is already comparable to the CIGO$+$space-based network configuration shown in Fig.~2. For $ f > 0.1\,\mathrm{Hz} $, TCIGO surpasses not only CIGO but also all existing or planned space-based interferometers. Overall, within our target frequency band, TCIGO improves the angular resolution by a factor of 5 compared to CIGO.
	
	\section*{Discussion}
	We investigated the angular resolution of LISA, TianQin, and CIGO for monochromatic gravitational waves in the frequency band 0.1--10\, Hz. The results show that CIGO exhibits excellent source-localization capability. For $f > 2.87$\, Hz, CIGO's angular resolution surpasses that of TianQin and LISA by several orders of magnitude, and the joint localization performance of the three-detector network is dominated by CIGO. Similar to TianQin, the normal vector to the CIGO detector plane is aligned with the lunar north pole and thus points toward the exact ecliptic longitude on the sky; in this direction, the signal-to-noise ratio is significantly reduced, resulting in reduced localization accuracy (larger $\Delta\Omega_s$) and weak amplitude/phase modulation. Consequently, in the sky maps, the region around ecliptic longitude $\phi_s = 5^\circ$ exhibits noticeably lower angular resolution than other regions. In contrast, LISA benefits from continuously rotating detector-plane orientation along its orbit, which provides amplitude modulation and yields more uniform sky coverage. In addition, we investigate the impact of the detector-noise formalism on angular-resolution modeling and show that CIGO would achieve better performance if low-frequency lunar seismic noise is further suppressed.
	
	To eliminate the angular-resolution blind spot caused by CIGO's fixed orientation, we further placed an additional detector at the bottom of the same lunar crater. Together with the original three CIGO detectors, this forms a regular tetrahedral configuration. The tetrahedral geometry introduces three additional oriented planes, yielding six new independent interferometric channels. The results show that, across the observed frequency band, the angular resolution of each new detector alone is slightly lower than that of the original CIGO triad. However, when combined with the original detectors, the overall localization accuracy improves by a factor of approximately 5. Moreover, because the detector orientations now provide complete sky coverage, the sky coverage blind spots previously present in CIGO are effectively eliminated.
	
	The lunar-based CIGO, particularly in its tetrahedral configuration, delivers outstanding localization performance for mid-frequency gravitational-wave sources in the 0.1--10\, Hz band. It thereby serves as a vital bridge between space-based detectors (LISA and TianQin), which dominate the low-frequency regime ($10^{-4}$--$10^{-1}$\, Hz), and ground-based observatories (LIGO, Virgo, KAGRA, Einstein Telescope, Cosmic Explorer), which cover the high-frequency band (10--$2\times10^{4}$\, Hz). This unique setup makes CIGO an indispensable component of gravitational-wave astronomy in the coming decades

	\section*{Methods}

	\subsection*{General GW Signal and Response in GR}
	In general relativity, a gravitational wave signal is described by the strain tensor
	\begin{equation}
		h_{ij}(t) = \sum_{A=+, \times} e_{ij}^A h_A(t),
		\label{eq:gw_signal}
	\end{equation}
	where $h_A(t)$ is the GW strain amplitude in the polarization mode $A$, and $e_{ij}^A$ is the corresponding polarization tensor. At the quadrupole leading order, the two polarizations of a monochromatic gravitational wave with frequency $f_0$ take the form:
	\begin{equation}
		\begin{split}
			h_+ &= \mathcal{A} \left[ 1 + (\vec{L} \cdot \hat{\omega})^2 \right] \exp(2\pi i f t + i \phi_0), \\
			h_\times &= 2 i \mathcal{A} (\vec{L} \cdot \hat{\omega}) \exp(2\pi i f t + i \phi_0),
		\end{split}
		\label{eq:monochromatic_polarizations}
	\end{equation}
	Here  $\mathcal{A} = 2 M_1 M_2 / (r d_L)$ denotes the amplitude, determined by the binary masses $M_1$,$M_2$, the distance $r$ between them and the and the luminosity distance $d_L$ between the GW source and the observer, $\vec{L}$ represents the unit vector for the binary's orbital angular momentum, and $\phi_0$ is the initial phase. 
	
	Then the output of the detector $\alpha$ is

	\begin{equation}
		s_\alpha(t) = \left[ \sum_A F^A_\alpha h_A(t) e^{i \phi_D(t)}\right] + n_{\alpha }(t),
		\label{eq:detector_output}
	\end{equation}
	
	where $A = +,\times$ denotes the plus and cross polarizations,
	$n_{\alpha }(t)$ is the instrumental noise, $e^{i \phi_D(t)}$ is the doppler phase modulation
	factor. 
	The response function ${F}^A_{\alpha }$ of detector $\alpha$ for the polarization mode $A$ can be written as
	\begin{equation}
		{F}^A_{\alpha }= \sum_{i,j} D^{ij}_\alpha e_{ij}^A,
		\label{eq:angular_response}
	\end{equation}
	where $D^{ij}_\alpha$ is the detector tensor for detector $\alpha$, and $e_{ij}^A$ is the GW polarization tensor. The detector tensor $D^{ij}$ represents the instrument's response to signals from specific directions. For a monochromatic gravitational wave with frequency $f$ propagating along $\hat{\omega}$, in an equal-arm Michelson laser interferometer with single round-trip light travel, $D^{ij}_\alpha$ is given by
	\begin{equation}
		D^{ij} = \frac{1}{2} \left[ \hat{u}^i \hat{u}^j T(f, \hat{\omega}) - \hat{v}^i \hat{v}^j T(f, \hat{\omega}) \right],
		\label{eq:detector_tensor}
	\end{equation}
	respectively. The transfer function $T(f, \hat{\omega})$ is given by \cite{blaut2011accuracy,hu2018fundamentals,cutler1998angular,zhang2021sky,cornish2001space}
	\begin{equation}
		\begin{split}
			T(f, \hat{u} \cdot \hat{\omega}) &= \frac{1}{2} \Biggl\{ \mathrm{sinc}\left[ \frac{f (1 - \hat{u} \cdot \hat{\omega})}{2 f^*} \right]  \times \exp\left[ -i\frac{f (3 + \hat{u} \cdot \hat{\omega})}{2 f^*} \right] \\
			&\quad + \mathrm{sinc}\left[ \frac{f (1 + \hat{u} \cdot \hat{\omega})}{2 f^*} \right]  \times \exp\left[ -i\frac{f (1 + \hat{u} \cdot \hat{\omega})}{2 f^*} \right] \Biggr\},
		\end{split}
	\end{equation}
	where $\hat{u}_\alpha$ and $\hat{v}_\alpha$ denote the unit vectors along the two arms of the detector $\alpha$, and $\hat{\omega}$ is the incident direction vector of the gravitational wave. $f^* = c / (2\pi L)$ is the transfer frequency, where $L$ is the arm length and $c$ is the speed of light. For CIGO, which primarily targets the 0.1--10\, Hz band, the GW's wavelength is much larger than the detector arm length (100\, km). This implies that the observing frequency $f \ll f^{*}$, so $\mathrm{sinc}(x) \to 1$ and the transfer function approaches unity: $T(f, \hat{u} \cdot \hat{\omega}) \to 1$. Consequently, in low frequency limit, the detector tensor $D^{ij}$ reduces to its long-wavelength, frequency-independent form:
	
	\begin{equation}
		D^{ij} = \frac{1}{2} \left( \hat{u}^i \hat{u}^j - \hat{v}^i \hat{v}^j \right).
	\end{equation}
	
	The modulation factor on the detector induced by the Earth-Sun and Moon-Earth motions
	$e^{i \phi^\alpha(t)}$, can be expressed as follows \cite{zhang2021sky}
	\begin{equation}
		e^{i \phi^\alpha(t)} = \exp\left(i\frac{2\pi  f}{c} R \cos\theta_s \cos\left(\frac{2\pi t}{T} - \phi_s - \phi_\alpha\right)\right),
		\label{eq:modulation_factor}
	\end{equation}
	where $f$ is the GW frequency, $R$ is the orbit radius of the detector center in the heliocentric-ecliptic coordinate system, $ (\theta_s, \phi_s)$ is the spherical coordinates of the gravitational-wave source on the celestial sphere, $\phi_\alpha$ is the ecliptic longitude of the detector $\alpha$ at t = 0. The detailed derivation of this formula is presented in Supplementary Information C. For CIGO, we can calculate using the equivalent radius by direct geometric projection:
	\begin{equation}
		R = R_{E} + \sin\theta_{m}·R_{M},
	\end{equation}
	where $R_{E}$ and $R_{M}$ are the radius of Earth and Moon orbit. 
	Due to the long duration and large scale of the CIGO mission, we often set $R_{E}$ to 1\, AU and the observation time $T$ to 1 year. $\theta_m$ is the orbital inclination of the lunar orbit plane in Fig. 8.
	Given the significant gap in their orders of magnitude, $R_{M}$ (0.00257 AU) can be neglected.

	\subsection*{ Lunar detector orbit and detector tensor}
	
	To calculate the detector tensor, it is necessary to determine the orbital equations of each station in CIGO. We adopt a heliocentric-ecliptic coordinate system (Fig. 8), with the x axis pointing toward the vernal equinox, the y axis completing the right-handed triad, and the z axis normal to the ecliptic plane. The n-th detector position can be decomposed as
	\begin{equation}
		\vec{r}_n(t) = \vec{r}_E(t) + \vec{r}_M(t) + \vec{r}'_n(t) + \vec{\epsilon}_n(t),
		\label{eq:detector_position}
	\end{equation}
	where $\vec{r}_E(t)$ represents the position vector of the Earth in the heliocenteric frame, $\vec{r}_M(t)$ represents the position vector of the Moon relative to Earth, and $\vec{r}'_n(t)$ represents the position vector of the detector $n$ relative to the Moon, where $n \in \{1,2,3\}$.$\vec{\epsilon}_n(t)$ is the high order perturbations correction to the orbits from the moon, the Earth, the Sun, other planets (mostly Jupiter), etc \cite{hu2018fundamentals}.
	
	The position vector of the $n$-th detector is given by
	\begin{equation}
		\begin{split}
			X_n(t) &= R \sin \theta_c \cos \alpha_n(t) + R_M \cos \alpha_M(t) + R_E \cos \alpha_E(t) \\
			&\quad + \frac{1}{2} e_1 R_M \left[ \cos(2\alpha_M(t)) - 3 \right] - \frac{3}{2} e_1^2 R_M \sin^2 \alpha_M(t) \cos \alpha_M(t) \\
			&\quad + \frac{1}{2} e_2 R_E \left[ \cos(2\alpha_E(t)) - 3 \right] - \frac{3}{2} e_2^2 R_E \sin^2 \alpha_E(t) \cos \alpha_E(t) \\
			&\quad + \mathcal{O}(3),
		\end{split}
		\label{eq:X_t}
	\end{equation}
	\begin{equation}
		\begin{split}
			Y_n(t) &=\left[  R \sin \theta_c \sin \alpha_n(t) + R_M \sin \alpha_M(t) + \frac{1}{2} e_1 R_M \sin(2\alpha_M(t)) \right. \\
			&\quad \left. + \frac{1}{2} e_1^2 R_M (3 \cos2\alpha_M(t) - 1) \sin \alpha_M(t) \right] \sin \theta_m \\
			&\quad + \left[ R \cos \theta_c + R_E \sin \alpha_E(t) + \frac{1}{2} e_2 R_E \sin2\alpha_E(t) \right. \\
			&\quad \left. + \frac{1}{2} e_2^2 R_E (3 \cos(2\alpha_E(t)) - 1) \sin \alpha_E(t) \right] \cos \theta_ + \mathcal{O}(3),
		\end{split}
		\label{eq:Y_t}
	\end{equation}
	\begin{equation}
		\begin{split}
			Z_n(t) &= R \sin \theta_m \cos \theta_c \\
			&\quad - \left[ R \sin \theta_c \sin \alpha_n(t) + R_M \sin \alpha_M(t) + \frac{1}{2} e_1 R_M \sin2\alpha_M(t) \right. \\
			&\quad \left. + \frac{1}{2} e_1^2 R_M (3 \cos2\alpha_M(t) - 1) \sin \alpha_M(t) \right] \cos \theta_m + \mathcal{O}(3),
		\end{split}
		\label{eq:Z_t}
	\end{equation}
	where $\mathcal{O}(3)$ denotes the higher-order infinitesimal to the eccentricity. Here, \(R \) is the radius of the moon; \( R_M \) and \( e_1 \) are the semi-major axis and eccentricity of the Moon's orbit around the Earth; \( R_E \) and \( e_2 \) are the semi-major axis and eccentricity of the Earth's orbit around the Sun, respectively. The angle \( \alpha_n(t) = 2\pi f_n t + \kappa_n \) is the angular displacement of the \( n \)-th detector in the detector-plane–lunar polar coordinate system after time \( t \), where \( \kappa_n = 2\pi (n-1)/3 \). The term \( \alpha_M(t) = 2\pi f_M t + \kappa_M \) represents the mean lunar longitude of the Moon in the geocentric–lunar coordinate system, where \( f_M \) is the modulation frequency due to the Moon's orbital motion and \( \kappa_M \) is the mean lunar longitude at \( t = 0 \). Similarly, \( \alpha_E(t) = 2\pi f_E t + \kappa_E \) denotes the mean ecliptic longitude of the Earth in the heliocentric–ecliptic coordinate system, where \( f_E \) is the orbital frequency of the Earth and \( \kappa_E \) is the mean ecliptic longitude measured from the vernal equinox at \( t = 0 \). $\sin\theta_c = {d}/{R}$, where \( d\) is the distance from the center of the detector to each detector. At the same time, $\theta_m$ is the angle between the lunar orbit plane and the ecliptic.
	The detailed derivation is provided in Supplementary Information A.
	
\begin{figure*}[t] 
	\centering
	\includegraphics[width=0.95\textwidth]{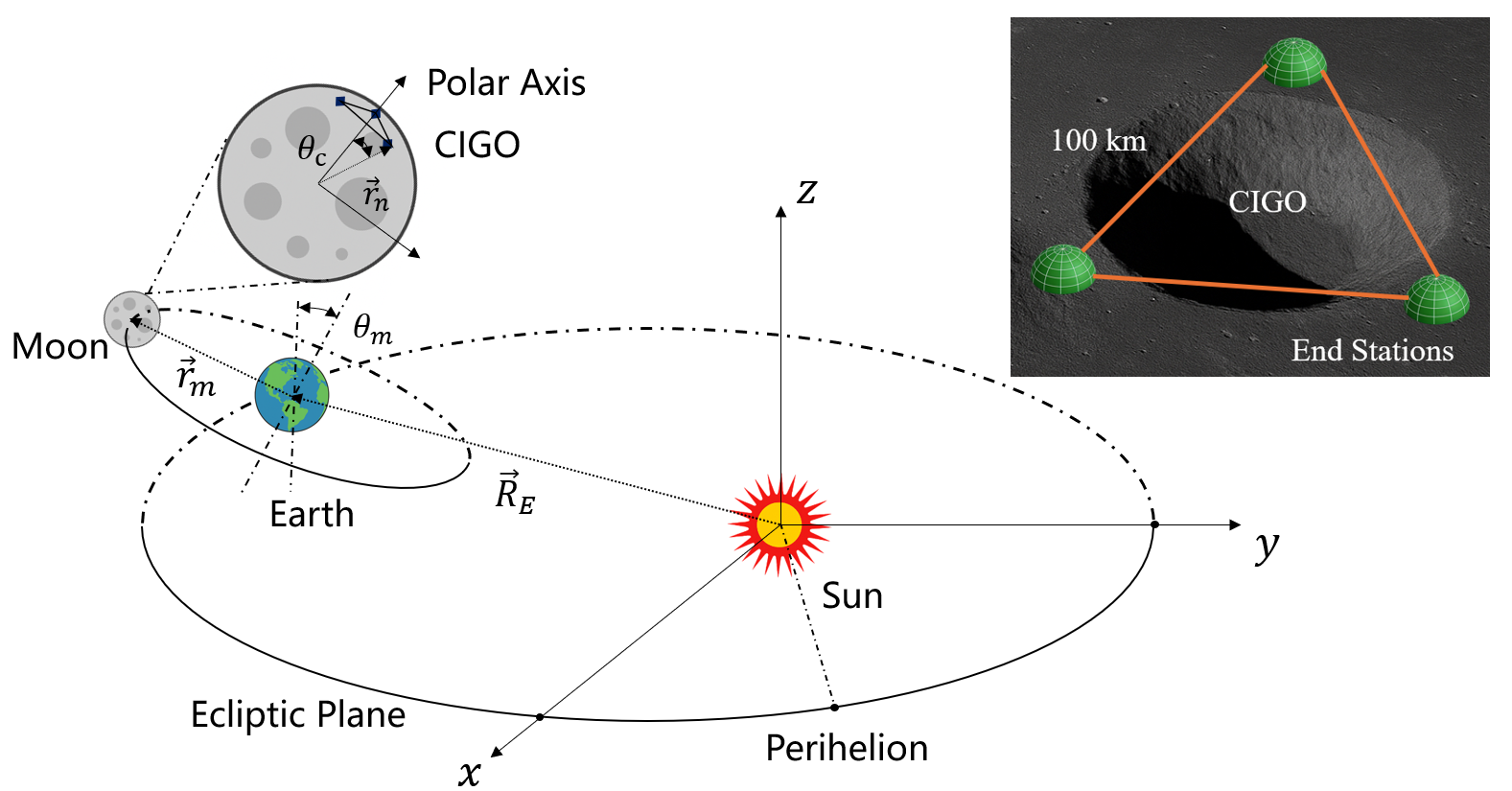} 
		\caption{Schematic diagram of CIGO's orbit and coordinate system.
			The detector is located at the Moon's north pole. The Moon orbits the Earth with inclination $\theta_m$ relative to the ecliptic plane, while the Earth--Moon system revolves around the Sun. 
			The inset shows the conceptual design of CIGO: three end stations (SC1--SC3) are placed on 
			the rim of a lunar crater, forming an approximately equilateral triangular configuration 
			with arm lengths of about $100\,\mathrm{km}$. Each station hosts a test mass and three laser 
			links forming the interferometric baselines.
		} 
		\label{fig:8}
	\end{figure*}

	\subsection*{Fisher Information Matrix}
	The monochromatic gravitational waves considered in this work can be probed by interferometers such as CIGO, as well as by other space-based detectors.
	We adopt seven parameters in our Fisher information matrix (FIM) analysis:
	\begin{equation}
		\theta = (\theta_s, \phi_s, \theta_L, \phi_L, \mathcal{A}, \phi_0, f),
	\end{equation}
	which are source direction $(\theta_s, \phi_s)$; the direction of the binary's orbital angular momentum $(\theta_L, \phi_L)$; The amplitude $\mathcal{A}$, the initial phase $\phi_0$, and frequency $f$. 
	
	Then the FIM takes the form as \cite{blaut2011accuracy,hu2018fundamentals,cutler1998angular,zhang2021sky,cornish2001space, zhang2022source}:
	\begin{equation}
		\begin{aligned}
			\Gamma_{ij} &= \sum_{\alpha} \left\langle \frac{\partial s_{\alpha}}{\partial \theta_i} \;\middle|\;\frac{\partial s_{\alpha}}{\partial \theta_j} \right\rangle \\
			&= \sum_{\alpha} \left[ \frac{2}{S_{n}(f_0)} \left( \int_{-\infty}^{\infty} {\partial _is_{\alpha}(t)} {\partial_j s^{*}_{\alpha}(t)} \, dt \right) \right],
		\end{aligned}
		\label{eq:fim_corrected}
	\end{equation}
	
	in which $\theta_i$ denotes the $i$-th parameter and ${\partial _is_{\alpha}}={\partial s_{\alpha}}/{\partial \theta_i} $. Accordingly, the corresponding covariance matrix is given by
	\begin{equation}
		\sigma_{ij} = \left\langle \Delta \theta_i \Delta \theta_j \right\rangle \approx (\Gamma^{-1})_{ij}.
		\label{eq:covariance_detailed}
	\end{equation}
	The $S_n(f)$ is the single-link noise power spectral density, which we have discussed in "Result” section.
	The angular uncertainty in sky localization is quantified as
	\begin{equation}
		\Delta \Omega_s = 2\pi \left| \cos \theta_s \right| \sqrt{ \sigma_{\theta_s\theta_s} \sigma_{\phi_s\phi_s} - \sigma_{\theta_s \phi_s}^2 },
		\label{eq:sky_uncertainty}
	\end{equation}
	with $\theta_s$ and $\phi_s$ denoting the source sky coordinates. We now have the basic setup for the sky-map resolution, and we have present numerical results in "Result” section.
	
	
\bibliographystyle{unsrt} 
\bibliography{reference.bib}       
	
\end{document}